# Unusual Fe-H bonding associated with oxygen vacancies at the (001) surface of $Fe_3O_4$

Fangyang Liu[1], Chen Chen[1], Hangwen Guo[1], Mohammad Saghayezhian[1], Gaomin Wang[1], Lina Chen[1], Wei Chen[2], Jiandi Zhang[1], E. W. Plummer[1]*

[1]Department of Physics and Astronomy, Louisiana State University, Baton Rouge, Louisiana 70803-4001, USA.

[2]Department of Physics, Harvard University, Cambridge, Massachusetts 02138, United States.

*Correspondence to:  wplummer@phys.lsu.edu

**Abstract**

An unusual Fe-H bonding rather than conventional OH bonding is identified at $Fe_3O_4$ (001) surface. This abnormal behavior is associated with the oxygen vacancies which exist on the surface region but also penetrate deep into the bulk $Fe_3O_4$. In contrast, OH bonding becomes preferential as generally expected on an ozone processed surface, which has appreciably less oxygen vacancies. Such bonding site selective behavior, depending on oxygen vacancy concentrations, is further confirmed with DFT calculations. The results demonstrate an opportunity for tuning the chemical properties of oxide surfaces or oxide clusters.

**Introduction**

The adsorption of hydrogen on oxides is a fundamental chemical reaction. This reaction is associated with many areas in science and technology, especially in catalysis, hydrogen storage, and environmental science. Tremendous effort has been devoted to understand the accompanying interactions at the atomic level, including studies of surfaces of bulk oxide crystals[1, 2] as well as oxide nanoparticles[3]. However, the presence of oxygen vacancies raises the complexity of the problem substantially. Oxygen vacancies have been reported to play a significant role in chemical reactions. For example, oxygen vacancies are demonstrated to assist the coupling and enolization of acetaldehyde on a $CeO_2$ (111) surface[4], and to enhance the catalytic activity of $MnO_2$[5, 6]. However, the microscopic origin of such vacancy-related chemical reactions is still under intensive investigation. The identification of oxygen vacancy locations and its role in the bonding to adsorbate atoms like H is crucial for understanding the nature of associated chemical interactions.

Although being a well-known magnetic material in the history [7], $Fe_3O_4$ also serves as a catalyst in today's chemical industry [8]. The partially occupied iron d-band makes its surface ideal for adsorption and dissociation of molecules [9]. $Fe_3O_4$ has a cubic inverse spinel lattice structure, consisting of two different layers (A and B). As shown in Fig. 1, the tetrahedral (A type) layer has center sites occupied by $Fe^{3+}$ cations while the octahedral (B type) layer has center sites of equal numbers of $Fe^{3+}$ and $Fe^{2+}$ cations. Oxygen anions reside on the corners of the tetrahedron and octahedron. Experimentally, the (001) surface is terminated with B layer ((1×1) symmetry in the bulk). The surface reconstructs, forming a $(\sqrt{2} \times \sqrt{2})R45°$ phase to remove the polarity of this surface [10]. Several models have been proposed to explain the properties of the reconstructed surface [10-13], but they all assume a defect free surface. Adsorption of H onto the (001) $Fe_3O_4$ surface removes the reconstruction [14], where chemical intuition and theoretical calculations [15] predict that the H bonds to O atoms at the surface. The bonding site has not been determined experimentally.

In this article, we report a striking observation that H exposure on the conventionally processed (CP) (001) surface of $Fe_3O_4$ results in mainly Fe-H rather than OH bonding (see Fig. 1). But on ozone processed (OP) surface, conventional OH bonding becomes dominant after H exposure. We argue that oxygen vacancies are the key to this abnormal H bonding preferentialism. First-principle calculations also illustrate that the high density of oxygen vacancies is the origin of this peculiar chemical behavior for CP magnetite. Furthermore, we demonstrate that the presence of O vacancies is not just localized at the surface of the oxide but can extend deep into the bulk, thus potentially impacting more than surface properties.

**Experimental Details**

All experiments were performed in ultrahigh vacuum (UHV) systems using natural magnetite single crystal samples purchased from SurfaceNet and cut in (100) direction. The resistivity vs. temperature measurements on these samples exhibit a Verwey transition at T ≈ 120 K. The (001) surfaces were prepared through cycles of sputtering (1keV $Ne^{++}$ 15min) and annealing (900K in vacuum 30min). The surface was then further annealed in $1\times10^{-6}$ Torr oxygen environment at 800K, which creates the CP surface. The OP surface was annealed in 1 mTorr 1% ozone environment at 800K after cycles of sputtering and annealing in vacuum. After both treatments, sharp $(\sqrt{2} \times \sqrt{2})R45°$ low energy electron diffraction (LEED) patterns were observed. Surface composition was examined by X-ray photoelectron spectroscopy (XPS) to check for surface contamination. Hydrogen adsorption was done by exposing the surfaces to atomic hydrogen at room temperature with a line of sight hot W filament in the presence of a partial pressure of $H_2$.

Low-energy ion scattering spectroscopy (LEIS) experiments were performed in a UHV system with base pressure $2\times10^{-10}$ Torr. A 1500eV $He^+$ beam with a ~200nA current was used to probe the $Fe_3O_4$ surfaces. The incident and scattering angle were both fixed at 50°. In these condictions, the sputtering rate is more than two hours per layer [16]. High-resolution electron energy loss spectroscopy (HREELS) data were collected by a LK-5000 spectrometer in a μ-metal shielded chamber with base pressure $5\times10^{-11}$ Torr. A 7 eV electron beam was tuned by two monochromaters, and then scattered off the sample surface. The energy loss of the scattered electron beam was analyzed with an instrumental resolution of ~2 meV. The electron analyzer was located in the specularly reflected direction, with an incident angle of 70°. This allows the incoming electrons to interact with the long range dipole field from the surface, which originates from the bulk dielectric properties.

Sample transfer between the HREELS and XPS chamber were accomplished with a vacuum suitcase, which consists of an ion pump, transfer arm, ion gauge and gate valve. The base pressure therein can be maintained at $5\times10^{-9}$ Torr.

**Results and Discussion**

**3.1 H adsorption on CP surface**

Parkinson *et al.* have observed bright protrusions on top of Fe atoms with STM and Fe valence reduction with XPS for H covered $Fe_3O_4$ (001) surface [14], but there has been no direct evidence for the location of the adsorbed H. After atomic hydrogen exposure, LEED pattern returns to unreconstructed $p(1\times1)$ pattern, as reported by Parkinson *et al.* [14] [see the inset of Fig. 2(A)-(B)]. Fig. 2 shows our measurements for the adsorption of atomic H on CP surface at room temperature, using LEED, LEIS and HREELS.

LEIS is an extremely surface sensitive technique used to characterize chemical and structural properties [17]. Usually an incident beam of inert gas ions is scattered back from the surface, losing energy and changing momentum [inset of Fig. 2(C)]. The backscattering can be modeled by a simple classical calculation involving the energy of the incident beam and the mass of the surface atoms [17]. Atomic H cannot be detected with this technique, but the adsorbed H can shadow the atoms below, referred as shadowing effect, giving an indication of the bonding configuration [18, 19]. Fig. 2(C) shows the LEIS spectra for the CP surface before and after H adsorption, using $He^+$ ions. Before hydrogen adsorption, the ratio of the O (at ~ 960 eV) to Fe (at ~ 1280 eV) peak areas is 47:100, which agrees with a previous report [20]. After hydrogen exposure, the Fe peak drops dramatically by ~ 40% while the O peak is only reduced by ~25%. This indicates H prefers the surface Fe site, though the reduction in the O signal is also significant, and will be shown to be related to the OH intensity in the XPS spectra.

In principle, HREELS is an ideal tool to determine the H bonding site by looking for the H stretching mode, whose energy will reflect the bonding configuration. The OH stretching vibrational mode usually has energy near 450 meV [21], while the energy for the Fe-H vibrational mode is ~110 meV for H on a bridge site of Fe(110) [22]. Unfortunately, the spectrum for ionic crystals is dominated by intense Fuchs-Kliewer (FK) vibrational modes [23], which are determined by bulk dielectric functions, and penetrate deep into the bulk, typically over 10 nm. Although FK modes are the result of the creation of a surface, they are not sensitive to surface conditions. The properties of FK phonon modes have been reviewed by Kress et al. [24]. These intense bulk-like modes make it very difficult to see the true surface vibrations [24, 25]. The lower panel of Fig. 2(D) displays the HREELS spectra before and after hydrogen adsorption on the CP surface. No peak in the energy range of 450 meV was observed for exposures of 10L to 1800L at both 77K and 300K. To observe the spectral region of the Fe-H mode, a Fourier Transform Deconvolution (FTD) procedure is performed. This method proposed by P.A. Cox in 1984 has been successfully applied to ZnO and $SrTiO_3$ HREELS spectra [26]. After FTD [Fig. 2(D) upper panel], a peak at 113 meV appears, indicative of a Fe-H bond. Thus, HREELS results confirm the picture of Fe-H bonding rather than OH bonding.

## 3.2 H adsorption on OP surface

Although the result that Fe-H bonding is more preferential compared to O-H bonding is surprising, similar phenomenon have been observed by LEIS on highly oxygen deficient $TiO_2$(001) surface [27], where only the Ti peak is decreased after hydrogen exposure (O peak remains unchanged). This prompted our belief that oxygen vacancies may play an important role in bonding site selectivity. Fig. 3 displays the results of H adsorption on OP surface, showing that H bonds to the O when the number of oxygen vacancies has been reduced. Similar to CP, a clean OP surface is also $(\sqrt{2} \times \sqrt{2})R45°$ reconstructed, showing that the surface symmetry doesn't change upon ozone processing. The OP surface changes to a $p(1\times1)$ structure after H exposure. However, the HREELS data in Fig. 3(A) clearly reveals the presence of an OH stretching mode at ~ 450 meV, which was not seen for H adsorption on the CP surface. This OH peak is broad, most likely due to inhomogeneous broadening associated with the multitude of possible sites on a surface with defects. The Fe-H vibrational mode (shown in Fig. 2(D)) is not observed for the OP surface. The XPS O 1s spectrum taken at emission angle $\theta = 81°$ (which is much more surface sensitive than for normal emission) in Fig. 3 (B) clearly shows a new high binding energy peak, which is due to the OH bond. Fitting the experimental curve gives an intensity distribution of 64% ± 1.4% for $O^{2-}$ at ~ 530.2 eV, 13% ± 3.8% in the $O_{vac}$ at ~ 531.2 eV, and 23% ± 3.5% in the OH peak at ~ 532.1 eV. The inset in Fig. 3(B) displays normal emission XPS and shows little change upon adsorption of H.

For comparison, Fig. 3(C) shows the XPS data of H adsorption on a CP surface. An appreciable reduction in the area of the O 1s spectral intensity (~ 10%) is observed when H is adsorbed. There is also a redistribution of the spectral weight, with the high binding energy "vacancy" peak ($O_{vac}$) increasing to ~ 44% compared to 35% for the clean CP surface. Normal emission XPS data [inset of Fig.3(C)] for the same conditions show almost no change induced by H adsorption on the CP surface. We speculate that H adsorption creates more oxygen vacancies at the surface, or that part of the spectral weight of the $O^{2-}$ peak was stolen by the formation of a small amount of OH bonds. We can fit the spectra taken with $\theta = 81°$ for the H-covered CP surface in Fig. 3(C) with the same weight (44%) in the $O_{vac}$ peak at ~ 531.4 eV as for the clean CP surface and get ~8% spectral weight in the OH peak at ~ 532.2 eV (Fig. 3(C)) as well as 48% in the $O^{2-}$ at ~ 530.4 eV. The OH spectral intensity is not seen in the more bulk sensitive normal emission [inset of Fig. 3(C)].

The completely different bonding sites upon H adsorption for the CP and OP surfaces are determined to be associated with oxygen vacancies. XPS O 1s core level spectra are shown in Fig. 4(A) and (B) for normal emission from clean CP and OP surfaces, respectively. The mean free path for the electrons excited from the O 1s core is 2.5 nm [28], meaning that normal emission probes 2.5 nm into the bulk, i.e. 10 AB layers in Fig. 1. The spectra in the insets of Fig. 4(A) and (B) are for the 81° grazing emission angle, which probes only the top AB layer. The O 1s core level XPS spectrum for the CP surface shown in Fig. 4(A) is much more asymmetric (with extra spectral weight at high binding energy side) [29] than the OP surface spectrum (Fig. 4(B)). Fan *et al.* fitted the spectra for a CP surface with two peaks, attributing the higher binding energy peak (~531.5 eV) to the presence of oxygen vacancies [30]. Fig. 4(A) shows the fitting to our spectra with two peaks: one at 531.4 eV, with 35%±1.1% of the total spectral intensity at the "vacancy" peak ($O_{vac}$), and the other at 530.4 eV for $O^{2-}$ with 65%±0.7% of the spectral intensity. The spectral weight of the "vacancy" peak is reduced to 18%±3.9% in OP surface spectra, showing that the relative oxygen vacancy concentration is appreciably reduced. There is no measurable difference in the ratio of the "vacancy" peak to the $O^{2-}$ peak with large emission angle for either CP or OP surfaces (see inset Fig.4 (A) and 4(B)). Thus O vacancies for both CP and OP surfaces are not localized at the surface, but penetrate deep into the bulk.

HREELS is routinely used to measure the vibrational motion of atoms and molecules on a surface [31], and it can also provide information on lattice conditions. The HREELS data shown in Fig. 2(D) & Fig. 3(A) for H adsorption on CP and OP surfaces emphatically illustrate different FK modes. The FK modes are sharper, more intense, and better defined on an OP surface than for CP surfaces. According to White and De Angelis [32], the 50 meV and 79 meV peaks are induced by the Fe-O antisymmetric stretch and bending mode in tetrahedral groups. As stated previously, adsorption of foreign gases on the surface has almost no effect on the FK modes [24, 25, 33], so the differences in the FK modes indicate more oxygen vacancies in CP surface regions, which in turn changes the physical properties deep into the bulk.

These conclusions are critically important for understanding the physical and chemical properties of the $Fe_3O_4$ (001) surface. To our knowledge, all previously reported experiments use CP to prepare the (001) surface of $Fe_3O_4$ single crystals. However, we show that CP samples are oxygen deficient in the selvedge (subsurface) region near the surface. Therefore, the results of previous studies need to be analyzed in the context of an oxygen deficient selvedge. It is also significant that neither CP nor OP samples have oxygen vacancies only at the surface. The changes in the FK modes and the angle dependence of the O 1s core level spectra in both cases indicate that the O vacancies penetrate deep into the bulk.

### 3.3 Desorption studies on H covered surfaces

The results shown above have described the changes induced by adsorption of atomic H for both CP and OP surfaces. To understand the mechanism of this abnormal Fe-H bonding on a CP surface, desorption experiments were performed. Fig. 5(A)-(C) display the evolution of LEED patterns after heating a CP sample to different temperatures. The patterns in Fig. 5(A) and (B) show the surface with a (1×1) structure, originally induced by H adsorption. The surface only returns to the "clean" $(\sqrt{2} \times \sqrt{2})R45°$ structure when heated to 700K [Fig.5(C)]. The LEIS spectra in Fig. 5(D) indicate that heating to 500K removes the H bonded to the Fe, as witnessed by the Fe peak returning to its initial intensity. However, the O peak in the LEIS spectra [see the inset of Fig. 5(D)] has not recovered all of the intensity lost by H adsorption, i.e. there is still missing O or O atoms are being shadowed by H, and the LEED pattern is still (1×1) (Fig. 5 (B)). If the sample is heated to 700K, 200K above the desorption temperature for Fe-H, the LEED pattern, LEIS, and XPS spectra return to the original CP surface condition. An obvious explanation is that the small amount of H bonded to O on CP surface is removed at a higher temperature than the H bonded to Fe.

Desorption of H from an OP surface was also studied using XPS. The primarily OH higher binding energy peak at 532 eV in the XPS O 1s spectrum is a signature of the H bond. Heating the sample to 570K removes about 70% of the intensity in this peak (dashed line in Fig. 5 (E) shows the OH fitted peak with ~7% of spectral intensity), but the sample must be heated to ~700K to completely remove the OH peak and return the LEED pattern to the "clean" $(\sqrt{2} \times \sqrt{2})R45°$ structure. This suggests H has a higher desorption temperature on an OP surface than on a CP surface, i.e. the OH bond on an OP surface is stronger than the Fe-H or OH bonds on a CP surface, which is intuitively correct, and reproduced by calculation. The reversible process of hydrogen adsorption on the CP and OP surfaces suggest that these two surfaces are both stable phases of B terminated $Fe_3O_4$, yet different oxygen vacancy concentrations lead to their distinct physical and chemical properties.

To illustrate the role of oxygen vacancies on the adsorption of atomic H, we have performed self-consistent DFT calculations for the binding of H as a function of the concentration of O vacancies on the surface. Density functional theory (DFT) calculations were performed using the Vienna ab initio simulation package (VASP)[34] with projector-augmented wave (PAW) potentials[34, 35]. The generalized gradient approximation (GGA) parameterized by Perdew-Burke-Ernzerhof (PBE)[36] for the exchange-correlation functional was used. The energy cutoff was 400 eV for the plane-wave basis sets. We used U = 5 eV and J = 1 eV for the Fe 3d orbitals, similar to the choices in previous studies of $Fe_3O_4$[37].

For a defect free surface we reproduce the results of Mulkaluri and Pentcheva [15] for H bonded to O with a binding energy of -1.31 eV. As the O vacancy concentration at the surface increases, the binding energy of H to O decreases while the H to Fe bond strength increase. There is a crossover at ~37% vacancy concentration, and at 50% vacancy concentration, the Fe-H bond is -1.05 eV compared to -0.42 eV for the OH bond. This picture is consistent with the above experiments, showing that Fe-H bonding is preferred on highly oxygen deficient (CP) surfaces, but small amounts of OH may also exist. These DFT calculations for a model defect density illustrate the importance on knowing the metadata associated with sample processing. Undoubtedly, the selvedge oxygen vacancies should be considered in a more elaborate calculation.

**Conclusion**

In summary, we discovered a hitherto unreported adsorption behavior of atomic hydrogen on $Fe_3O_4$ (001) surfaces. H bonds to Fe sites on an oxygen vacancy rich surface but to O sites on the less oxygen deficient surface. The stronger OH bonding will break at higher temperature when compared to Fe-H bonding during desorption. We demonstrated this unusual H bonding is associated with the presence of oxygen vacancies both experimentally and theoretically.

**References and Notes:**

[1] V.E. Henrich, The surfaces of metal oxides, Reports on Progress in Physics, 48 (1985) 1481.
[2] J.L. Rowsell, O.M. Yaghi, Effects of functionalization, catenation, and variation of the metal oxide and organic linking units on the low-pressure hydrogen adsorption properties of metal-organic frameworks, Journal of the American Chemical Society, 128 (2006) 1304-1315.
[3] G.V. Buxton, C.L. Greenstock, W.P. Helman, A.B. Ross, Critical review of rate constants for reactions of hydrated electrons, hydrogen atoms and hydroxyl radicals (· OH/· O− in aqueous solution, Journal of physical and chemical reference data, 17 (1988) 513-886.
[4] F.C. Calaza, Y. Xu, D.R. Mullins, S.H. Overbury, Oxygen vacancy-assisted coupling and enolization of acetaldehyde on CeO2 (111), Journal of the American Chemical Society, 134 (2012) 18034-18045.
[5] L. Li, Z. Wei, S. Chen, X. Qi, W. Ding, M. Xia, R. Li, K. Xiong, Z. Deng, Y. Gao, A comparative DFT study of the catalytic activity of MnO 2 (211) and (2-2-1) surfaces for an oxygen reduction reaction, Chemical Physics Letters, 539 (2012) 89-93.
[6] D.A. Tompsett, S.C. Parker, M.S. Islam, Rutile (β-) MnO2 surfaces and vacancy formation for high electrochemical and catalytic performance, Journal of the American Chemical Society, 136 (2014) 1418-1426.
[7] R. Hartwell, Markets, technology, and the structure of enterprise in the development of the eleventh-century Chinese iron and steel industry, The Journal of Economic History, 26 (1966) 29-58.
[8] M. Munoz, Z.M. de Pedro, J.A. Casas, J.J. Rodriguez, Preparation of magnetite-based catalysts and their application in heterogeneous Fenton oxidation–A review, Applied Catalysis B: Environmental, 176 (2015) 249-265.
[9] R. Schlögl, Ammonia Synthesis, Wiley Online Library, 2008.
[10] S. Chambers, S. Joyce, Surface termination, composition and reconstruction of $Fe_3O_4$(001) and γ-$Fe_2O_3$(001), Surface science, 420 (1999) 111-122.
[11] F. Voogt, T. Fujii, P. Smulders, L. Niesen, M. James, T. Hibma, $NO_2$-assisted molecular-beam epitaxy of $Fe_3O_4$, $Fe_{3-δ}O_4$, and γ-$Fe_2O_3$ thin films on MgO(100), Physical Review B, 60 (1999) 11193.
[12] J. Rustad, E. Wasserman, A. Felmy, A molecular dynamics investigation of surface reconstruction on magnetite (001), Surface science, 432 (1999) L583-L588.
[13] R. Bliem, E. McDermott, P. Ferstl, M. Setvin, O. Gamba, J. Pavelec, M. Schneider, M. Schmid, U. Diebold, P. Blaha, Subsurface cation vacancy stabilization of the magnetite (001) surface, Science, 346 (2014) 1215-1218.
[14] G.S. Parkinson, N. Mulakaluri, Y. Losovyj, P. Jacobson, R. Pentcheva, U. Diebold, Semiconductor–half metal transition at the $Fe_3O_4$(001) surface upon hydrogen adsorption, Physical Review B, 82 (2010).
[15] N. Mulakaluri, R. Pentcheva, Hydrogen Adsorption and Site-Selective Reduction of the $Fe_3O_4$(001) Surface: Insights From First Principles, The Journal of Physical Chemistry C, 116 (2012) 16447-16453.
[16] G. Betz, G.K. Wehner, R. Behrisch, Sputtering by particle bombardment II, Topics in Applied Physics, 52 (1983) 11.
[17] J.W. Rabalais, Principles and applications of ion scattering spectrometry: surface chemical and structural analysis, Wiley New York, 2003.

[18] R. Bastasz, T. Felter, W. Ellis, Low-energy He[+] scattering from deuterium adsorbed on Pd (110), Physical review letters, 63 (1989) 558.
[19] H. Brongersma, M. Draxler, M. Deridder, P. Bauer, Surface composition analysis by low-energy ion scattering, Surface Science Reports, 62 (2007) 63-109.
[20] B. Stanka, W. Hebenstreit, U. Diebold, S. Chambers, Surface reconstruction of $Fe_3O_4$(001), Surface science, 448 (2000) 49-63.
[21] S. Liu, S. Wang, W. Li, J. Guo, Q. Guo, Water Dissociation on Magnetite (001) Films, The Journal of Physical Chemistry C, 117 (2013) 14070-14074.
[22] A. Baro, W. Erley, The chemisorption of hydrogen on a (110) iron crystal studied by vibrational spectroscopy (EELS), Surface Science Letters, 112 (1981) L759-L764.
[23] R. Fuchs, K. Kliewer, Optical Modes of Vibration in an Ionic Crystal Slab, Physical Review, 140 (1965) A2076-A2088.
[24] F.W. de Wette, Study of Surface Phonons by the Slab Method, in: W. Kress, F. de Wette (Eds.) Surface Phonons, Springer Berlin Heidelberg, 1991, pp. 67-109.
[25] K. Wolter, D. Scarano, J. Fritsch, H. Kuhlenbeck, A. Zecchina, H.-J. Freund, Observation of a localized surface phonon on an oxide surface, Chemical Physics Letters, 320 (2000) 206-211.
[26] P. Cox, W. Flavell, A. Williams, R. Egdell, Application of Fourier transform techniques to deconvolution of HREEL spectra, Surface Science, 152 (1985) 784-790.
[27] J.M. Pan, B. Maschhoff, U. Diebold, T. Madey, Interaction of water, oxygen, and hydrogen with $TiO_2$(110) surfaces having different defect densities, Journal of Vacuum Science & Technology A, 10 (1992) 2470-2476.
[28] C. Powell, A. Jablonski, Evaluation of calculated and measured electron inelastic mean free paths near solid surfaces, Journal of Physical and Chemical Reference Data, 28 (1999) 19-62.
[29] S.A. Chambers, Fe 2p Core-Level Spectra for Pure, Epitaxial α-Fe2O3(0001), γ-Fe2O3(001), and Fe3O4(001), Surface Science Spectra, 5 (1998) 219.
[30] J.C. Fan, J.B. Goodenough, X-ray photoemission spectroscopy studies of Sn-doped indium-oxide films, Journal of Applied Physics, 48 (1977) 3524-3531.
[31] H. Ibach, D.L. Mills, Electron energy loss spectroscopy and surface vibrations, Academic press, 2013.
[32] W. White, B. DeAngelis, Interpretation of the vibrational spectra of spinels, Spectrochimica Acta Part A: Molecular Spectroscopy, 23 (1967) 985-995.
[33] K. D'amico, F. McFeely, E.I. Solomon, High resolution electron energy loss vibrational studies of carbon monoxide coordination to the (10. hivin. 10) surface of zinc oxide, Journal of the American Chemical Society, 105 (1983) 6380-6383.
[34] G. Kresse, J. Furthmüller, Efficient iterative schemes for ab initio total-energy calculations using a plane-wave basis set, Physical Review B, 54 (1996) 11169.
[35] P.E. Blöchl, Projector augmented-wave method, Physical Review B, 50 (1994) 17953.
[36] J.P. Perdew, K. Burke, M. Ernzerhof, Generalized gradient approximation made simple, Physical review letters, 77 (1996) 3865.
[37] R. Pentcheva, F. Wendler, H.L. Meyerheim, W. Moritz, N. Jedrecy, M. Scheffler, Jahn-Teller Stabilization of a "Polar" Metal Oxide Surface: $Fe_3O_4$(001), Physical Review Letters, 94 (2005).

**Acknowledgments:** We would like to thank Rongying Jin, Phillip Sprunger, and Richard Kurtz for useful discussions.   This material is based upon work supported as part of the Center for Atomic Level Catalyst Design, an Energy Frontier Research Center funded by the U.S. Department of Energy, Office of Science, Office of Basic Energy Sciences under Award Number DE-SC0001058. This project was partially funded by research support from the Office of Research and Economic Development at LSU

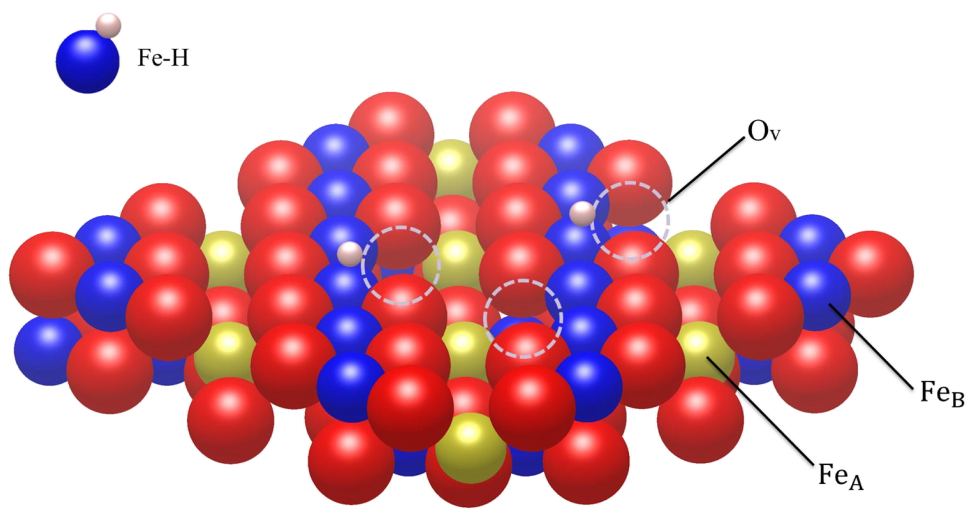

**Fig. 1.** Structural model for the (001) surface truncated form $Fe_3O_4$ crystal with the presence of oxygen vacancies and hydrogen adsorption.

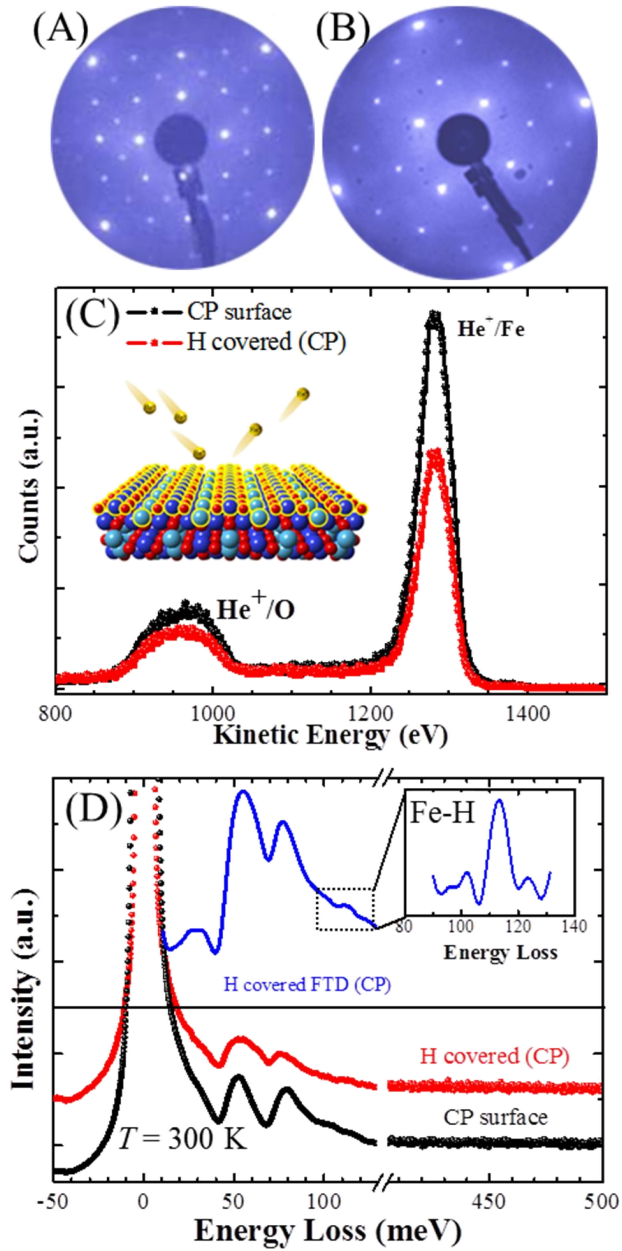

**Fig. 2.** LEED pattern of (A) clean and (B) H-covered CP surface, respectively. (C) LEIS spectra (1500 eV He$^+$) measured from both surfaces. Inset shows the shadowing effect of LEIS, where the highlighted first layer atoms will shadow the atoms underneath from being scattered with incident ions. (D) HREELS spectra, which show no OH stretching mode at ~ 450 meV for H-covered surface. After Fourier Transform Deconvolution (FTD), a small peak appears at ~113 meV (upper panel), the inset shows the peak with background subtracted.

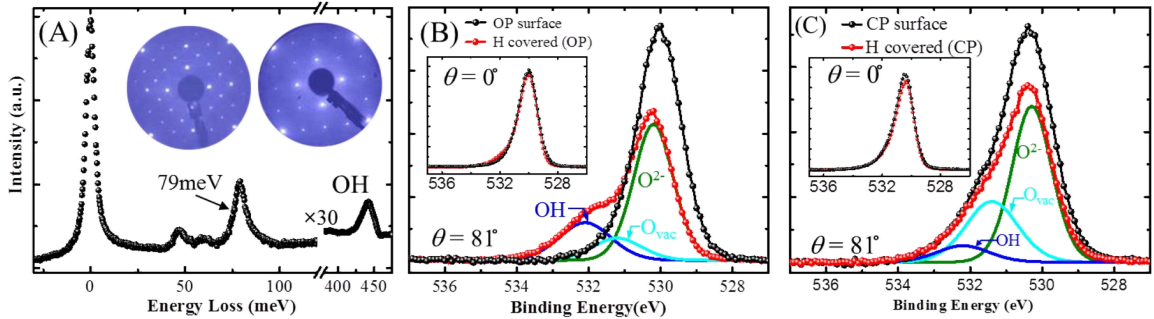

**Fig. 3.** (A) HREELS spectrum for H covered OP surface; Inset: LEED pattern of (*left*) OP and (*right*) H-covered surface (90 eV), respectively. XPS O 1s spectra for clean (black) and H-covered (red) (B) OP and (C) CP surface taken at emission angle $\theta = 81°$, respectively. The inset shows the corresponding spectra taken at normal emission ($\theta = 0°$). The spectra of H covered surfaces are fitted with three components, $O^{2-}$, O vacancy ($O_{vac}$)- and OH-related peaks. All the measurements are made at room temperature.

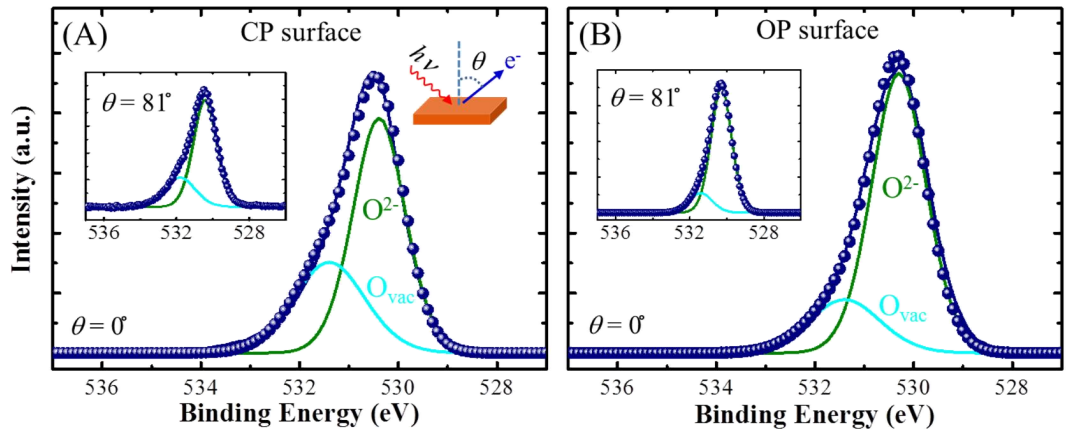

**Fig. 4.** XPS O 1s spectrum of the clean (A) CP and (B) OP surface measured at normal emission ($\theta = 0°$). The insets show the corresponding spectra taken at emission angle $\theta = 81°$ [see the schematic of the XPS setup in the inset of (A)]. All the spectra are fitted with $O^{2-}$ (530.4 eV) and O vacancy related peak (531.4 eV).

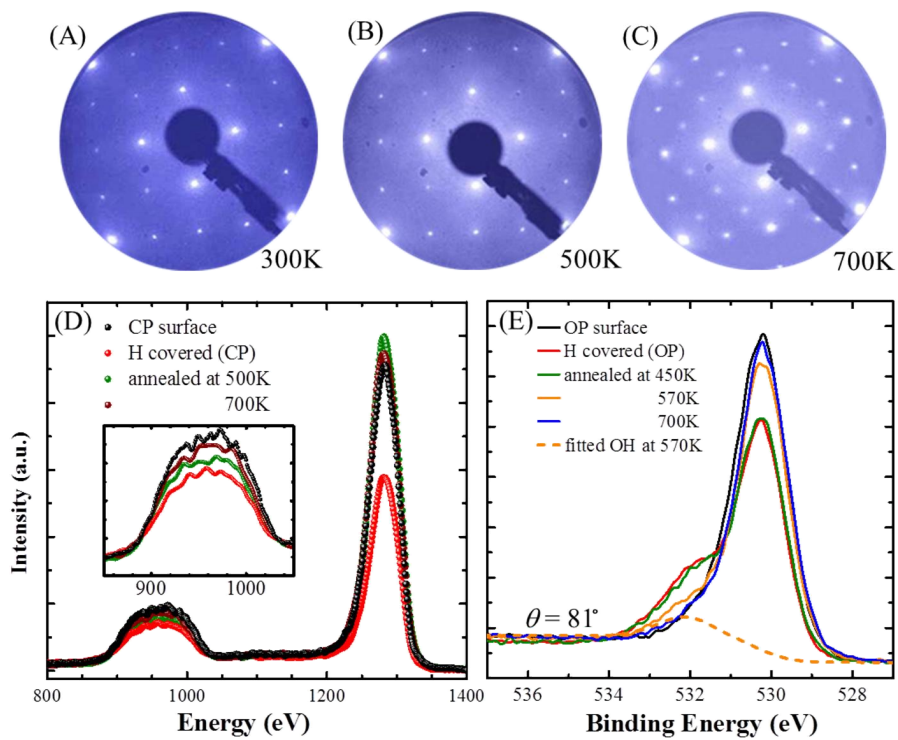

**Fig. 5.** Desorption process: (A)-(C) LEED pattern of H-covered CP surface after heating sample to 300K, 500K and 700K, respectively; (D) LEIS spectra of H covered CP surface and annealed at different temperatures. Fe peak recovered at 500K while O peak recovered at 700K; (E) XPS O 1s spectrum of H covered OP surface and annealed at different temperatures. The hydroxyl-related peak at ~532 eV is significantly reduced after heating to 570K.